\documentclass[nofootinbib,a4paper,aps,pra,showpacs,preprintnumbers,twocolumn]{revtex4-1}
\usepackage{cmap}
\usepackage{graphics,graphicx,epsfig}
\usepackage{epstopdf}
\usepackage[centertags]{amsmath}
\usepackage{amsfonts}
\usepackage{amssymb}
\usepackage[cp1251]{inputenc}
\usepackage[english]{babel}
\inputencoding{cp1251}
\usepackage{graphicx}
\usepackage{amsthm,color}
\usepackage{euscript}
\graphicspath{{pics/}}
\DeclareGraphicsExtensions{.pdf,.png,.jpg,.eps}
\begin{document}
\def\BY{\begin{eqnarray}}
\def\EY{\end{eqnarray}}
\def\L{\label}
\def\nn{\nonumber}
\def\ds{\displaystyle}
\def\o{\overline}
\def\({\left (}
\def\){\right )}
\def\[{\left [}
\def\]{\right]}
\def\<{\langle}
\def\>{\rangle}
\def\h{\hat}
\def\h{\hat}

\title{Squeezed supermodes and cluster states based on modes with orbital angular momentum}
\author{Vashukevich E.A., Losev A.S., Golubeva T.Y., Golubev Y.M.}
\date{\today}
\address{Saint-Petersburg State University, St Petersburg, 199034, Russia}
\begin{abstract}
In this paper, we discuss the possibility of building a linear cluster state based on modes with a certain orbital angular momentum (OAM). We show that in the system under consideration a field with a rich mode structure is generated in the cavity. We also analyze the conditions under which an infinite system of Heisenberg – Langevin equations describing the dynamics of intracavity fields can be shortened and solved analytically. To analyze the genuine number of quantum degrees of freedom, we use the supermodes technique. This approach allows us to build the most entangled cluster state.
\end{abstract}
\maketitle
\section{Introduction}
Interest in the problems associated with the generation of cluster states of light is caused by the
widespread use of such states in the protocols of one-way quantum computations. Although initially the
principles of cluster states generation have been formulated for discrete variables \cite{Raussendorf2001},
in recent years multipartite entangled quantum systems in continuous variables have attracted  more and more
interest \cite{Menicucci2006,Braunstein1999,Zhang2006}. An important feature of such systems is the ability
to generate a large-scale cluster entangled states \cite{Pfister2012,Furus2014}. From this point of view,
there is an obvious interest in using light with an orbital angular momentum \cite{Woerdman1992}  (OAM) as a
resource for generating cluster states, since the quantum number associated with the OAM can take any
integer values, which makes it possible to increase the number of degrees of freedom of the system
indefinitely.

It should be noted that the presence of a large number of correlated modes in the system does not yet
guarantee an equal scale of cluster state. A vivid example here is the radiation of a SPOPO - synchronously
pumped optical parametric oscillator. Despite the fact that SPOPO radiation has an extremely wide spectrum
(about $10^6$  frequency modes), the number of genuine quantum degrees of freedom of such a system is much
smaller (about $100$) \cite{TrepsSPOPO2012}. Thus, it is insufficiently just to reveal the interacting modes
for analyzing the size of a cluster state, but it is necessary to find the appropriate measurement basis.

In this paper, we will consider the process of spontaneous parametric down-conversion in the cavity with a
nonlinear parametric crystal placed inside. The cavity is pumped by an external field, which possessing an
OAM. A similar problem has already been considered by the authors in  \cite{chinese}, however, our approach
differs both in terms of the theoretical justification of the methods used and in the analysis of the
results obtained. Applying the techniques developed in the papers \cite{Treps2014,Treps2012,Treps2006}, we
show that the dimension of the cluster state is less than the number of initial modes. Also, we analyze the
optimal parameters of cavity pumping for cluster state generation.
\section{Optical parametrical oscillator below the threshold}
\subsection{Theoretical model}
In this paper, we will consider the following system: a crystal with quadratic nonlinear
susceptibility is placed in a spherical or self-imaging cavity (Fig.\ref{Fig.1}), whose eigenmodes are the
full set of Laguerre-Gaussian modes \cite{selfimFabre2011, Benlloch2009}. The system is pumped by two
spatial Laguerre-Gaussian (LG) modes $U^{LG}_{0,1} $ and $U^{LG}_ {0,-1}$  propagating in z-direction:
\BY
&&U_{p,l}^{LG}\propto\left(\frac{\rho \sqrt2}{w(z)} \right)^{|l|} \EuScript
{L}_{p}^{|l|}\left(\frac{2\rho^2}{w^2(z)}\right)\times\nn\\
&&\times \exp{\{\frac{-\rho^2}{w^2(z)}\}}\exp{\{il\phi\}},
\EY
where $z,\rho,\phi$ are cylindrical coordinates, indices $l$ and $p$ are integers that
define the transverse profile of the mode, $w(z)$ is the waist width of the transverse field distribution
inside the cavity, $\EuScript{L}_{p}^{|l|}$ is the associated Laguerre polynomial. The Laguerre-Gaussian
functions with different indices are orthogonal to each other over the whole space and normalized to
unity:
\BY
\int d \rho d\phi dz \;U_{p,l}^{LG}U_{p^\prime,l^\prime}^{LG*}=\delta_{p,p^\prime}\delta_{l,l^\prime}.
\EY

It has been shown \cite{Woerdman1992} that LG beams carry a certain orbital angular
momentum, which is determined by the phase factor $\exp{\{il\phi\}}$. The azimuth number $l$ is an
eigenvalue of the OAM operator, the index $l$ indicates the projection of the mode's orbital angular
momentum. In our case, we set the angular momentum of one pump wave to be $l_{pump}=1$, and of the other one to be $l_{pump}=-1$, both waves propagate at the frequency $\omega_{pump}$. The
choice of the pump structure is due to our further needs: we want to provide the conditions for the
generation of the cluster state of the field. We discuss possible cavity excitation schemes in section
\ref{2b}. The radial index $p$ define the mode's spatial transverse profile. For small values of the orbital
angular indices $l$  we can suggest with high accuracy that the value of the $p$ does not change in the
process of parametric down-conversion \cite{miatto2011, miatto2012}. Thus, we can further assume that the
pump is carried out by beams with a radial index $p=0$, which is equal for all waves involved in the
process.
\begin{figure}[h!]
\includegraphics[scale=0.68]{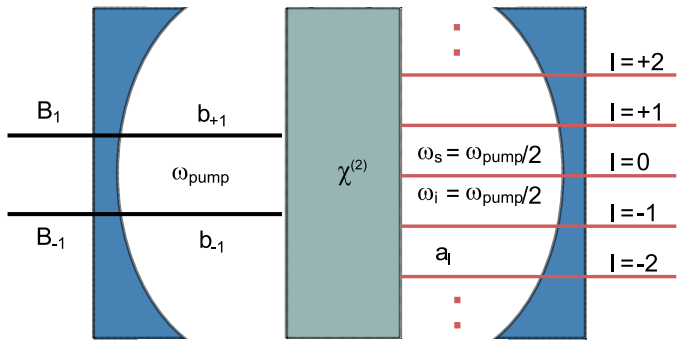}
\caption{
The schematic draw of the system under consideration: a crystal with quadratic nonlinearity, which ensures type-I phase synchronism, is placed in a cavity with
spherical mirrors. The pump consists of two spatial Laguerre-Gaussian
modes with OAM equal 1 and -1. A multimode field is generated in the resonator with different values of OAM but at
the same frequency.}
\L{Fig.1}
\end{figure}

The cavity supports both pump modes $\h b_1, \h b_{-1}$, and down-converted modes  $\h
a_l$. The crystal provides type-I phase matching. Since, as is known \cite{Lassen2009}, the process of
parametric signal conversion occurs with the conservation of the orbital moment, a field with an abundant
mode structure along the orbital angular momentum is generated in the cavity. We consider the conditions of
ideal phase matching: $\omega_{pump}=2\omega_i=2\omega_s,\;\;\vec{k}_{pump}=\vec{k}_s+\vec{k}_i$.

The interaction Hamiltonian can be written as follows:
\BY
&& \h H_I=i\hbar\sum\limits_{l}(\chi_{{}_{l,1-l}}\h b^{}_1\h a^\dag_{l} \h a^\dag_{1-l}+ \nn\\
&&+\chi_{{}_{l,-1-l}}\h b^{}_{-1}\h a^\dag_{l} \h a^\dag_{-1-l}) + H.c.,\L{ham}
\EY
where $\h b^{}_{\pm1}$ denotes photon annihilation operators in pump modes, $\h
a^\dag_{l}\;\;(l = 0, \pm1, \pm2, ...)$ are photon creation operators in signal and idler modes. These operators obey the following canonical commutation relations:
\BY
[\h a^{}_{l}, \h a^\dag_{l^\prime}]=\delta_{l, l^\prime},\;[\h b^{}_{\pm1}, \h b^\dag_{\pm1}]=1.
\EY

The effective coupling parameters $\chi_{{}_{l, \pm1-l}}$ are proportional to the overlap
integral between the pump, signal and idler modes. Their properties and impact on the characteristics of the system will be discussed below in subsection B. As is well known, frequency degeneration entails the generation of squeezed states of light. However,  such an effect is observed only for modes that do not possess an orbital angular momentum.
The appearance of an additional degree of freedom, such as OAM, leads to an OAM non-degenerate process.
Since the conservation law imposes a condition only on the total angular momentum of the waves involved in
the process, in this case, the total phase of the signal and idler waves is fixed: $\phi_{pump} = \phi_i +
\phi_s$, while the phases of each wave are remaining arbitrary. This means that in the process of generation
both squeezing and entanglement between the modes is formed.

\subsection{Heisenberg--Langevin equations below the oscillation threshold}\L{2b}
Considering the system below the threshold, we must neglect the process of pump exhaustion.
Therefore, we should take the pump modes remain constant and treat them classically.
Mathematically, this boils down to replacing the operators $\h b_{\pm1}$ with c-number quantities
$B_{\pm1}$. The Heisenberg-Langevin equations for signal and idler modes inside the cavity can be written as
follows:
\BY
\left\{
\begin{array}{l}
\dot{\h a}_{0} = -\gamma_0 \h a_{0} + 2 \chi_{0,1} ( B^{}_{1} \h a^\dag_{1}+  B^{}_{-1} \h a^\dag_{-1})  +\h
f_{0}\\
\dot{\h a}_{1} = -\gamma_1 \h a_{1} + 2 \chi_{0,1}  B^{}_{1} \h a^\dag_{0} +  2 \chi_{1,2}  B^{}_{-1} \h
a^\dag_{-2} +\h f_{1}\\
\dot{\h a}_{-1} = -\gamma_{-1} \h a_{-1} + 2 \chi_{0,1}  B^{}_{-1} \h a^\dag_{0} +  2 \chi_{1,2}  B^{}_{1}
\h a^\dag_{2} +\h f_{-1}\\
\ldots\\
\dot{\h a}_{k} = -\gamma_k \h a_{k} + 2 \sum\limits_{i=\pm1}\chi_{k,i-k}  B^{}_{i} \h a^\dag_{i-k} +\h
f_{k}\\
\dot{\h a}_{-k} = -\gamma_{-k} \h a_{-k} + 2 \sum\limits_{i=\pm1}\chi_{-k,i+k}  B^{}_{i} \h a^\dag_{i+k} +\h
f_{-k}\\
\ldots
\end{array}
\right.\L{eqn}
\EY
\begin{figure}[h!]
\includegraphics[scale=0.66]{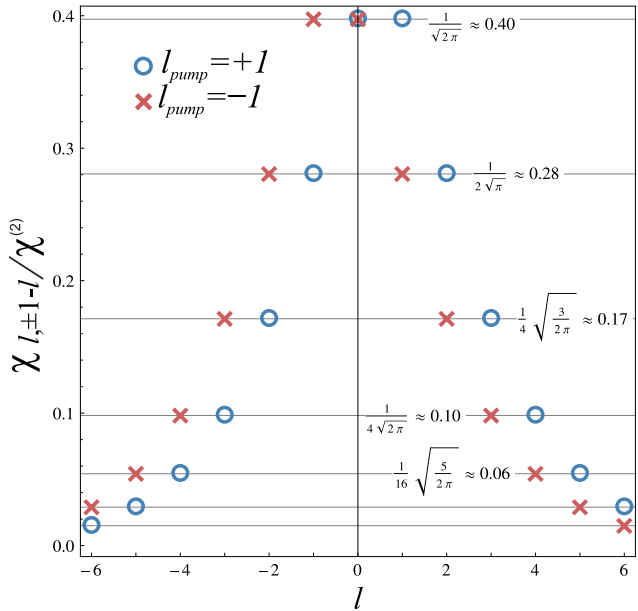}
\caption{The dependence of the coupling parameters normalized to the crystal quadratic
susceptibility $\chi^{(2)}$ on the orbital angular momentum of the signal mode, with OAM of pump  $l_{pump} =
1 $ (blue) and $ l_{pump} = - 1 $ (red). Here we calculate all the constants for the following parameters: a
thin crystal is located in the $z = 0$ plane, the ratio of the waist width of the signal mode to the waist
width of the pump mode $r=w_s(0)/w_p(0)=\sqrt{2}$.}
\L{Fig.2}
\end{figure}
It can be noted that the intracavity fields dynamics is described by an infinite number of
differential equations that mesh with each other. To solve them, we need to “shorten” the system of
equations by excluding all modes starting with a certain number $k$ from consideration. In \cite{chinese}, the
reason for shortening the system of equations was the decrease of the effective coupling constants
$\chi_{{}_{l,\pm1-l}}$ with increasing of index $l$. This treatment, however, is not sufficiently
substantiated, since a decrease in the coefficients of a system of differential equations does not yet
ensure the smallness of the corresponding variables. Moreover, it can be seen from Fig. \ref{Fig.2} that the
coefficients, that neighbouring in the equations, have the same order (with the selected calculation
parameters).
Each next coefficient is less than the previous one no more than
twice. Thus, in each particular equation, there
is no reason to neglect the third addent in comparison with the second one, as done in the paper
\cite{chinese}. It should be noted that a direct calculation confirms these preliminary conclusions: if we
compare the two sets of solutions obtained by solving a system of $k$ equations and $k + 1$ equations, it is
clear that such solutions are close only far from the oscillation threshold. However, as is well known, precisely the near-threshold region seems the most interesting for quantum effects observation. When
approaching the oscillation threshold, additional efforts are needed to ensure that a limited number of
modes can be considered.

It is appropriate to discuss here the choice of calculation parameters used in Fig.\ref{Fig.2}, which we will keep further.

Since our goal is to build a cluster state, we need to ensure a high efficiency of the
parametric down-conversion not for a single mode, but at the same time for several modes with different OAM.
The numerical calculation shows that the dependence of the coupling constants on the ratio of the waist widths has a
peculiar value for $r=\sqrt{2}$ (see Fig.\ref{Fig.2b}).

 At this point, the coupling parameters
$\chi^{}_{l_{pump},l,l_{pump}-l}$ for several modes achieve their maximum, i.e the down conversion process goes
more efficiently than with other values of this ratio. As shown in the paper \cite{miatto2011}, with such a
choice of $r$, the parametric down-conversion process to the modes with the index $p=0$ will be most efficient. So, we limited our consideration here to only the case of $p=0$. The values of the
$\chi^{}_{l_{pump},l,l_{pump}-l}$ obtained with this choice of parameters matches with those calculated in
\cite{chinese2}.
\begin{figure}[h]
\includegraphics[scale=0.7]{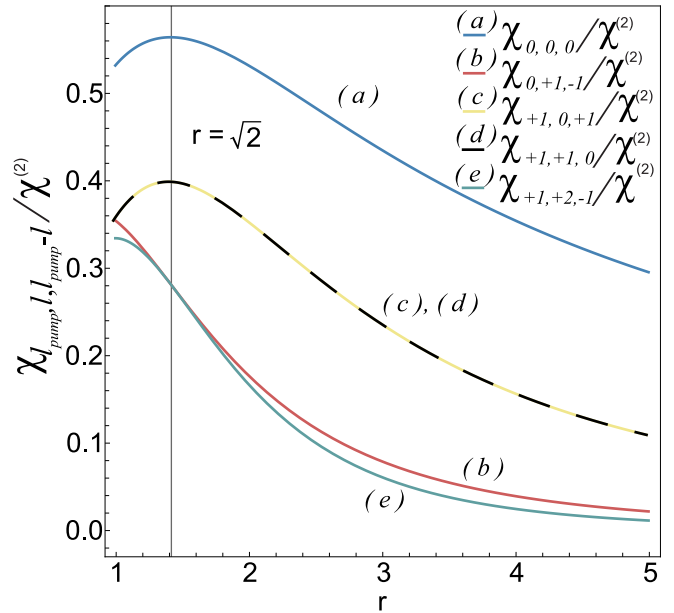}
\caption{The dependence of the normalized coupling parameters on the ratio of the waist width of the signal mode to the waist width of the pump mode $r=w_s(0)/w_p(0)$. Here the first index indicates the OAM of the pump mode.}
\L{Fig.2b}
\end{figure}
\subsection{Limitations on the number of modes considered}
The system of differential Eqs. (\ref{eqn}) can be shortened by changing the experimental conditions of a generation, for example, artificially providing a fast relaxation of the mode with the number $k$.

Hereinafter, we would like to limit our consideration to only five senior mods $a_i, i=0,\pm1,\pm2$. To do this, we assume that the decay rate of $a_{\pm3}$ modes is much greater than all the other constants in the system: $\gamma_{3}\gg\gamma_{i}, i=0,\pm1,\pm2$. There are several experimental methods based on the usage of special holograms that make it possible to select modes with a specific OAM \cite{Willner2014,Romanato2016}. For simplicity, the other relaxation constants are set equal to each other, $\gamma_{i}\equiv\gamma_{}, i\neq
3$. In this case, the system of differential Eqs. (\ref{eqn}) can be divided into two independent subsystems.

Since an analysis of the correlation properties of the light can be carried out based only on information about normally ordered means (and how they are related to ordinary means), we can simplify the problem and go from the system of Eqs. (\ref{eqn}) to c-number equations. C-number equations differ from Eqs. (\ref{eqn}) by the lack of operator designations and by corrected correlation functions for stochastic noise sources (see Appendix A). The standard procedure for such a conversion is described in detail in \cite{Davidovich}.

Further, for simplicity, we assume that the amplitudes of the pump waves $B_1$ and $B_{-1}$ are real numbers equal to each other. Given these conditions, the Eqs. (\ref{eqn}) can be rewritten for the quadrature components of the Glauber amplitudes $a$ and $a^*$:
$x_j=\frac{1}{2}(a_j+a_j^*),\;\;\;y_j=\frac{1}{2i}(a_j-a_j^*),\;\;\; f_j=f^\prime_j+if_j^{\prime\prime}\;\;\;,j=0,\pm1,\pm2$.

\BY
\left\{
\begin{array}{l}
\dot x_0=-\gamma x_0 + \gamma\mu(x_1 + x_{-1})  + f_{0}^\prime\\
\dot y_0=-\gamma y_0 - \gamma\mu( y_1 + y_{-1}) + f_{0}^{\prime\prime}\\
\dot x_{\pm1}=-\gamma x_{\pm1} +\gamma\mu x_1 +\gamma\xi x_{\mp2}  + f_{\pm1}^\prime\\
\dot y_{\pm1}=-\gamma x_{\pm1} -\gamma\mu x_1 -\gamma\xi x_{\mp2}  + f_{\pm1}^{\prime\prime}\\
\dot x_{\pm2}=-\gamma x_{\pm2} + \gamma\xi x_{\mp1}  + f_{\pm2}^\prime\\
\dot y_{\pm2}=-\gamma y_{\pm2} - \gamma\xi y_{\mp1}  + f_{\pm2}^{\prime\prime}
\end{array}
\right.
\L{qd}
\EY

Here, the quadratures of Langevin noise sources $f_{i}^\prime, f_{i}^{\prime \prime}\; (i=0, \pm1, \pm2)$ are defined by the following non-zero correlators (see Appendix A for more details):
\BY
\langle f_{0}^\prime(t) f_{\pm1}^\prime(t^\prime)\rangle = \frac{1}{2}\gamma\mu\delta(t-t^\prime)=
-\langle f_{0}^{\prime\prime}(t) f_{\pm1}^{\prime\prime}(t^\prime)\rangle,\;\;\\
\langle  f^\prime_{{\pm1}}(t), f^\prime_{{\mp2}}(t^\prime)\rangle=\frac{1}{2}\gamma\xi\delta(t-t^\prime)=
-\langle  f^{\prime\prime}_{{\pm1}}(t), f^{\prime\prime}_{{\mp2}}(t^\prime)\rangle.\nonumber
\EY

The equations  above are written using dimensionless pump parameters $\mu=\displaystyle\frac{2\chi_{0,1}B_1}{\gamma}$ and
$\xi=\displaystyle\frac{2\chi_{1,2}B_1}{\gamma}$. The pump parameters are nonnegative real numbers, their physical meaning and limits of variation will be discussed further in Section III.
\section{Solution of the equations and analysis of quantum degrees of freedom}
The simplest way to solve Eqs. (6) is to apply a Fourier transform that turns differential equations into algebraic equations and allows us to calculate the spectrum of the fluctuations of quadrature components. Performing a Fourier transform and considering a special case of zero frequency leads us to the following system of equations (the superscript $0$ denotes the Fourier image taken at zero frequency, the lower indices still indicate the mode number)
\BY
\left\{
\begin{array}{l}
\displaystyle\binom{x^0_{0}}{ y^0_{0}}=\pm\mu\binom{x^0_{1}+x^0_{-1}}{y^0_{1}+y^0_{-1}}+
\frac{1}{\gamma}\binom{f^{\prime0}_{0}}{f^{\prime\prime0}_{0}}\\
\\
\displaystyle\binom{x^0_{\pm1}}{ y^0_{\pm1}}=\pm\mu\binom{x^0_{0}}{y^0_{0}}\pm\xi\binom{x^0_{\mp2}}{y^0_{\mp2}}+
\frac{1}{\gamma}\binom{f^{\prime0}_{\pm1}}{f^{\prime\prime0}_{\pm1}}\\
\\
\displaystyle\binom{x^0_{\pm2}}{ y^0_{\pm2}}=\pm\xi\binom{x^0_{\mp1}}{y^0_{\mp1}}+
\frac{1}{\gamma}\binom{f^{\prime0}_{\pm2}}{f^{\prime\prime0}_{\pm2}}.
\end{array}
\right.\L{ft}
\EY
The system of Heisenberg--Langevin Eqs. (5), as well as the equations for the Fourier components (9), indicate that different modes do not evolve independently and “hint” the presence of entanglement shown in Fig.\ref{Fig.3}. Further calculation confirms this assumption.
\begin{figure}[!h]
\includegraphics[scale=0.6]{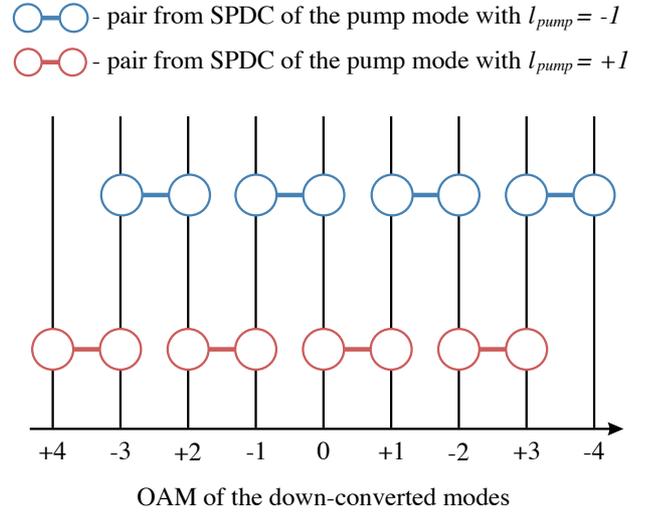}
\caption{The schematic of the entanglement between modes with different OAM.}
\L{Fig.3}
\end{figure}

Since the calculations in the initial basis are rather complicated, for a detailed analysis of the quantum properties of the system, we turn to the basis of summarized and differential modes:
\BY
&&\widetilde{ x}_0=x_0^0\nn,\;\;\;\widetilde{ x}_1= \frac{1}{\sqrt{2}}(x^0_{1}+x^0_{-1}),\;\;\;\widetilde{x}_2= \frac{1}{\sqrt{2}}(x^0_{2}+x^0_{-2}),\nn\\
&&\widetilde{ x}^-_1= \frac{1}{\sqrt{2}}(x^0_{1}-x^0_{-1}),\;\;\widetilde{x}^-_2= \frac{1}{\sqrt{2}}(x^0_{2}-x^0_{-2}).
\EY

Such a choice of variables allows us to obtain solutions of the Eqs. (\ref{ft}) and calculate the power spectra of fluctuations of the quadrature components and the cross-correlations (see Appendix B).
The calculation shows that there are correlations of the fluctuations in the $x$-quadrature and anti-correlations in the $y$-quadrature for the following modes:
\BY
&&\langle \delta \widetilde{y}_0\delta \widetilde{y}_{1}\rangle=-\langle \delta \widetilde{x}_0\delta \widetilde{x}_{1}\rangle=-\frac{\mu(1+ 2\mu^2 + \xi^2)}{\sqrt{2}\gamma(1-\xi^2-2\mu^2)^2},\nn\\
&&\langle \delta \widetilde{y}_{1}\delta \widetilde{y}_{2}\rangle=-\langle \delta \widetilde{x}_{1}\delta \widetilde{x}_{2}\rangle=-\frac{\xi(1+ 2\mu^2 + \xi^2)}{2\gamma(1-\xi^2-2\mu^2)^2},\nn\\
&&\langle \delta \widetilde{y}_0\delta \widetilde{y}_{2}\rangle=\langle \delta \widetilde{x}_0\delta \widetilde{x}_{2}\rangle=\frac{\sqrt{2}\mu\xi}{\gamma(1-\xi^2-2\mu^2)^2},\nn\\
&&\langle \delta \widetilde{y}^-_{1}\delta \widetilde{y}^-_{2}\rangle=-\langle \delta \widetilde{x}^-_{1}\delta \widetilde{x}^-_{2}\rangle=\frac{\xi(1+ \xi^2)}{2\gamma(1-\xi^2)^2}.
\label{quad}\EY
Getting back to the basis of the initial modes, one can verify the presence of quantum entanglement shown in Fig.\ref{Fig.3}.

Since we consider the below-threshold regime, the mean values
 of the intracavity field operators equal zero. All analysis was carried out in terms of the Glauber representation, so the variances of the Fourier transforms of the quadrature components, as well as their cross-correlation functions, coincide with the normally ordered means of the corresponding operators. Connection intracavity normally ordered means with mean values outside the cavity is determined by the well-known input-output relation:
\BY
\binom{\<|\delta \h X_i|^2\>}{\<|\delta \h Y_i|^2\>} = \frac{1}{4} + 2\gamma\binom{\<:|\delta \h x^0_i|^2:\>}{\<:|\delta
\h y^0_i|^2:\>},\;i=0,\pm1,\pm2\L{inout}.\;\;\;
\EY

Despite a simplification of analytical expressions as well as the presence of interesting correlations in the basis of summarized and differential modes, this basis, like the initial one, cannot be considered appropriate to reveal the genuine number of quantum modes of the system. Further, we show that the number of degrees of freedom, suitable for constructing a "good" $\;$ cluster state, is not equal to the number of initial modes considered.
\subsection{Supermodes of an optical parametric oscillator}
To find the optimal basis and analyze the number of quantum degrees of freedom, we use the technique developed in \cite{Treps2014}. As shown in this work, the field out of the optical parametric oscillator can be described either as an entangled state in the basis of individual modes with certain frequencies (in our case -- with a certain OAM) or as a set of uncorrelated squeezed states in the basis of the eigenvectors of the coupling matrix, attended in the interaction Hamiltonian of parametric down-conversion.
Let us rewrite interaction Hamiltonian (3) in the form:
\BY
\h H_I={i\hbar}\sum\limits_{i,s}M_{i,s}\h a^\dag_{i} \h a^\dag_{s}  + H.c.,\;\;\;\;i+s=\pm1.
\EY
The matrix elements $M_{i, s}$ here govern the coupling strength between two modes $\h a_{i} $ and $ \h a_{s}$. Matrix elements  $M_{i, s}$ are expressed in terms of eigenvectors and eigenvalues of the matrix $M$ as follows:
\BY
M_{i,s}= \sum\limits_{n}\lambda_n m_{n,i}m_{n,s},
\EY
where $\lambda_n$ are eigenvalues and $m_{n,i}$ are i-th element of a n-th eigenvector of the coupling matrix $M$.

Since M is a Hermitian matrix, it can be diagonalized. Let us define a new set of mods as a linear combination of the initial ones and derive the Hamiltonian in the new basis:
\BY
&& \h s^\dag_n = \sum\limits_{i}m_{n,i}\h a^\dag_{i},\nn\\
&&\h H_I=i\hbar\sum\limits_{n}\Lambda_n(\h s^\dag_{n})^2  + H.c.
\EY
Following the notation in [10] we will call the modes (14) as \textit{supermodes}.

One can note that the Hamiltonian, rewritten through the supermodes $\h s_n$, is the Hamiltonian of several separate degenerate processes occurring independently. Then supermodes manifest a quadrature squeezing imposed by degenerate process \cite{Treps2012, Treps2006}. The spectrum of $\{\Lambda_n\} $ determines the number of noncorrelated degrees of freedom and indicates the degree of squeezing in the modes $ \h s_n$.

Unlike the authors \cite{Fabre2010}, who use the supermodes technique to analyze infinite-dimensional matrices and therefore are forced to rely on numerical analysis, we are able to construct explicit analytical solutions. Based on the model presented above and limited to considering only five modes, we can write the matrix $M$ as follows:
\BY
M=\frac{\gamma}{4}\begin{pmatrix}
0&&\mu&&\mu&&0&&0\\
\mu&&0&&0&&0&&\xi\\
\mu&&0&&0&&\xi&&0\\
0&&0&&\xi&&0&&0\\
0&&\xi&&0&&0&&0\\
\end{pmatrix}.
\EY
Thus we can easily find eigenvalues and define the set of supermodes through initial modes:
\BY
&&\Lambda_1=0,\Lambda_2=-\frac{\gamma\xi}{4}=-\Lambda_3,\Lambda_4=-\frac{\gamma\sqrt{2\mu^2+\xi^2}}{4}=-\Lambda_5,\nn\\
&&\h s_1=-\frac{\xi}{\mu}\h a_0 + \h a_2 +\h a_{-2}\L{s1},\\
&&\h s_2=-\h a_1+\h a_{-1}-\h a_2+\h a_{-2},\\
&&\h s_3=\h a_1-\h a_{-1}-\h a_2+\h a_{-2},\\
&&\h s_4=\frac{2\mu}{\xi} \h a_0 -\frac{\sqrt{2\mu^2+\xi^2}}{\xi}(\h a_1+\h a_{-1})+\h a_2+\h a_{-2},\\
&&\h s_5= \frac{2\mu}{\xi}\h a_0 +\frac{\sqrt{2\mu^2+\xi^2}}{\xi}(\h a_1+\h a_{-1})+\h a_2+\h a_{-2}\L{s5}.
\EY
\subsection{Power of the fluctuations of the supermodes quadrature components}
In the previous section, we discussed the intracavity properties of the field in the supermode basis. Taking the explicit expressions for the correlators of the quadrature components of the initial modes (see Appendix B) and consider their connection with the supermodes (\ref{s1})-(\ref{s5}), we derive the spectral powers of the quadrature fluctuations of the supermodes $\h S_j = \h X^S_j + i \h Y^S_j \;(j = 1,2, ...)$ outside the cavity. Each of the supermodes we preliminarily normalize by unity and take into account the standard relations (\ref{inout}) between the extracavity and intracavity values:
\BY
&&\h s_j=\h x^s_j+i\h y^s_j, \<|\delta \h X^S_j|^2\>=\frac{1}{4} + 2\gamma \<:\h x^s_j:\>,\\
&&\<|\delta\h X^S_1|^2\>=\frac{1}{4},\\
&&\<|\delta\h X^S_2|^2\>=\frac{(\xi-1)^2}{4(\xi+1)^2}=\<|\delta\h Y^S_3|^2\>,\\
&&\<|\delta\h X^S_4|^2\>=\frac{(\sqrt{2\mu^2+\xi^2} -1)^2}{4(\sqrt{2\mu^2+\xi^2}+1)^2}=\<|\delta\h Y^S_5|^2\>.
\EY
It is clear from the expressions above that the positive eigenvalues of $\Lambda_3, \Lambda_5 $ correspond to supermodes squeezed in the $y$-quadrature, negative $\Lambda_2,\Lambda_4 $ -- in the $x$-quadrature. The $\h S_1$ mode is not squeezed, since there are no quadrature correlations between $\h a_0 $ and $\h a_{\pm2} $  (Appendix B), which is confirmed by the zero eigenvalue $\Lambda_0$. The power of the quadrature fluctuations in this mode do not depend on the pump parameters, and it can be said that the mode $ \h S_1$ is in a vacuum state.

Taking into account the selection of genuine quantum degrees of freedom for the further construction of the cluster state of multimode radiation, it is obvious that the use of a vacuum mode can only worsen the correlation of the system. Thus, we have shown that despite the presence of five quantum-correlated source modes, the genuine number of quantum degrees of freedom of this system is less by one.
\subsection{Parametric oscillation threshold and squeezing in the supermodes}
To determine the limits of applicability of the solutions obtained and the achievable ratio of squeezing, recall that the problem was solved in the approximation of inexhaustible pump, which means that the linear Eqs. (\ref{eqn}) are correct only for small fluctuations of the number of photons in the initial modes. As the fields in the cavity increase, the process of parametric up-conversion begins to play a significant role, and we can no longer assume that the pump amplitude remains constant in time.

From the solution of the Eqs. (\ref{ft}), it can be noted that the quadrature fluctuations in the Laguerre-Gaussian modes begin to increase indefinitely under the following condition:
\BY
2\mu^2+\xi^2\to1.\L{thr}
\EY
Thus, the limit expression (\ref{thr}) determines the threshold of parametric generation.

The degree of quadrature squeezing in supermodes obviously depends on the value of the pump parameters. Vacuum noise in the modes $\h S_4$ and $\h S_5$ are completely suppressed when approaching the generation threshold $2\mu^2+\xi^2=1$, and squeezing in these modes can be considered perfect. However, the modes $\h S_2$ and $\h S_3$ become perfectly squeezed only in the limit $\xi^2=1$. However this condition cannot be reached since even at $\xi^2>1-2\mu^2$ the system cannot be described by Eqs. (\ref{eqn}), and the problem should be considered in terms of the above-threshold regime. Hence, the perfect squeezing in these modes is unattainable.

Let us estimate the maximum degree of squeezing, taking into account the expression of the pump parameters through the constants $\chi_{l,\pm1-l}$. Restricting ourselves to the selected parameters of the Laguerre-Gaussian beams for the signal, idler, and pump modes (see Fig.\ref{Fig.2}), we can specify the following connection between two parameters:
\BY
\mu^2=2\xi^2.\label{pp}
\EY

Authors of \cite{miatto2011} assert that such a choice of beam parameters is optimal for the most efficient generation of a field with the mode structure under consideration. The condition for below-threshold field generation in the cavity is then rewritten as $\mu<\sqrt\frac{2}{5}$.

Summarizing all the above, we can conclude that when the pump parameter tends to the threshold, the squeezing in two supermodes will tend to the perfect, one supermode will be in a vacuum state, and the other two will be squeezed imperfectly:
\BY
&&\<|\delta \h X^S_4|^2\>=\<|\delta \h Y^S_5|^2\>\xrightarrow[\mu\to\sqrt{\frac{2}{5}}]{}0,\nn\\
&&\L{SQ}\<|\delta \h X^S_1|^2\>=\frac{1}{4},\\
&&\<|\delta \h X^S_2|^2\>=\<|\delta \h Y^S_3|^2\>\xrightarrow[\mu\to\sqrt{\frac{2}{5}}]{}\frac{(7 - 3 \sqrt{5})}{8}\nn.\L{SQ}
\EY

\section{Building of the cluster state of light}
We will construct a linear cluster state from supermodes, the properties of which we discussed in the previous section. The used supermodes $\h S_2-\h S_5$, described by variables ${\h X^S_i,\h Y^S_i}, \; i=2,3,4,5$, are  squeezed in $X$- and $Y$-quadrature alternately and independent from each other. In order to obtain a cluster state, we need to entangle the modes with each other in a certain way. The process of such entanglement can be come down to a unitary transformation over the original set of oscillators:
\BY
\begin{pmatrix}
\h Q_1 + i \h P_1\\
\h Q_2 + i \h P_2\\
\h Q_3 + i \h P_3\\
\h Q_4 + i \h P_4\\
\end{pmatrix} =U \begin{pmatrix}
\h X^S_3 + i \h Y^S_3\\
\h X^S_4 + i \h Y^S_4\\
\h X^S_5 + i \h Y^S_5\\
\h X^S_2 + i \h Y^S_2\\
\end{pmatrix},
\EY
where $ U $ is the transformation matrix, $ \hat{Q}_j $ and $\hat{P}_j $ are the quadrature components operators that describe the $j$ -th node of the cluster state (Fig.\ref{Fig.4}).

According to \cite{korolev2018}, there are certain requirements for the degree of quadrature squeezing of the mode used to build the cluster state. These requirements vary depending on the number of "neighbours" of the corresponding nodes of the cluster state. Based on this, we chose such an order of arrangement of the modes, so that the more squeezed modes were in the center, that is, they had two neighbours, and less squeezed --- on the edges.

\begin{figure}[h!]
\includegraphics[scale=0.58]{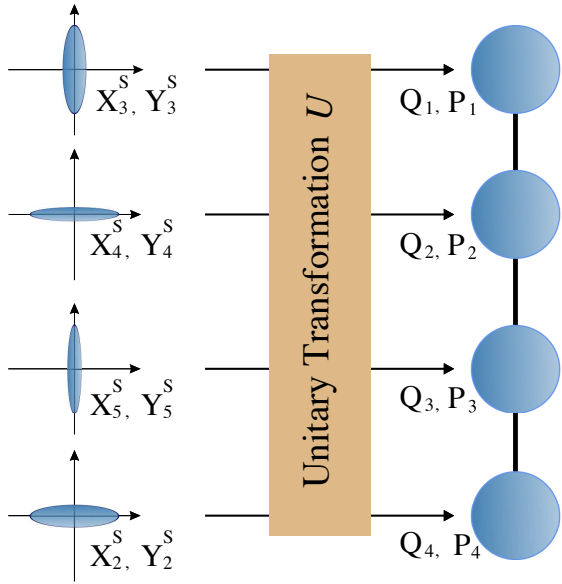}
\caption{Tranformation from squeezed supermodes to linear cluster state. Supermodes are quadrature squeezed alternately in $X$- and $Y$-quadrature, while the modes $\h S_3,\h S_2$ are squeezed imperfectly.}
\L{Fig.4}
\end{figure}

To determine the elements of the $ U $ matrix, as well as to analyze the resulting multipartite entangled state, we need to consider a mathematical description of this state. The cluster state can be described by a weighted undirected graph, whose nodes are treated as the modes of the physical system, described by pairs of canonical variables $\{\hat Q_i, \hat P_i \} $, edges - quantum entanglement between modes, and a set of edge weights sets the adjacency matrix $V$ of this graph.

To describe the quantum properties of the resulting state in continuous variables, it is common to use a set of nullifiers $\h N_i$, expressed in terms of quadratures of cluster nodes as follows:
\BY
\h N_i=\h P_i - \sum\limits_{j=1}^nV_{ij}\h Q_j,
\EY
where $ V_ {ij} $ are elements of the adjacency matrix.

Now we should define the adjacency matrix taking into account the cluster structure. In our case, for linear four-node cluster state (see Fig.\ref{Fig.4}) it is given in the form:
\BY
V=\begin{pmatrix}
0&&1&&0&&0\\
1&&0&&1&&0\\
0&&1&&0&&1\\
0&&0&&1&&0\\
\end{pmatrix},
\EY
that provide the following nullifiers:
\BY
&\h N_1=\h P_1-\h Q_2,\\
&\h N_2=\h P_2-\h Q_1-\h Q_3,\\
&\h N_3=\h P_3-\h Q_2-\h Q_4,\\
&\h N_4=\h P_4-\h Q_3.
\EY

According to \cite{ferrini2014}, the transformation matrix $U$ can be found through the adjacency matrix  $ V $ of the graph:
\BY
U=(I + iV)(1+V^2)^{-\frac{1}{2}}A,
\EY
where $I$ is the identity matrix, $A$ is any orthogonal matrix. It was shown that although the choice of the $A$ matrix affects the type of transformation over squeezed modes, it nevertheless does not change the structure of nullifiers and quantum properties of the resulting state.

Then, taking for simplicity $A$ equals identity matrix\footnotemark, we can express the nullifiers of the cluster state through the quadrature components of the supermodes:
\footnotetext{With this choice of the matrix $ A $, the transformation $ U $ looks as follows:
$$
\sqrt{\frac{2}{5}}\begin{pmatrix}
 \sqrt{1+\frac{1}{\sqrt{5}}} && -\sqrt{\frac{1}{2}+\frac{1}{\sqrt{5}}} && -\sqrt{\frac{1}{2}-\frac{1}{\sqrt{5}}} &&  \sqrt{\frac{1}{2}-\frac{1}{\sqrt{5}}} \\
 i \sqrt{\frac{1}{2}+\frac{1}{\sqrt{5}}} && -i \sqrt{\frac{1}{2}+\frac{1}{\sqrt{5}}} && i \sqrt{1-\frac{1}{\sqrt{5}}} &&-i\sqrt{\frac{1}{2}-\frac{1}{\sqrt{5}}} \\
 -\sqrt{\frac{1}{2}-\frac{1}{\sqrt{5}}} && -\sqrt{1-\frac{1}{\sqrt{5}}} && \sqrt{\frac{1}{2}+\frac{1}{\sqrt{5}}} &&  -\sqrt{\frac{1}{2}+\frac{1}{\sqrt{5}}} \\
 -i\sqrt{\frac{1}{2}-\frac{1}{\sqrt{5}}} && -i\sqrt{\frac{1}{2}-\frac{1}{\sqrt{5}}} && i \sqrt{\frac{1}{2}+\frac{1}{\sqrt{5}}} && i \sqrt{1+\frac{1}{\sqrt{5}}} \\
\end{pmatrix}$$}
\BY
&&\h N_1=\sqrt{1+ \frac{2}{\sqrt{5}}} \h Y^S_3 +\sqrt{1- \frac{2}{\sqrt{5}}} \h Y^S_5,\\
&&\h N_2=- \sqrt{2+ \frac{2}{\sqrt{5}}} \h X^S_4-\sqrt{1- \frac{2}{\sqrt{5}}} \h X^S_2,\\
&&\h N_3=\sqrt{1- \frac{2}{\sqrt{5}}} \h Y^S_3+ \sqrt{2+ \frac{2}{\sqrt{5}}} \h Y^S_5,\\
&&\h N_4=-\sqrt{1-\frac{2 }{\sqrt{5}}} \h X^S_4-\sqrt{1+ \frac{2}{\sqrt{5}}} \h X^S_2.
\EY
As can be seen from Eqs. (36)-(39), the nullifiers depend only on the squeezed quadratures, which guarantees the smallness of their variances. However, even near the generation threshold, the spectral powers of the fluctuations of the nullifiers do not vanish, since the expression for each nullifier includes the quadrature components of the non-perfect squeezed supermodes $\h S_3,\h S_2 $.

To show that the obtained state is exactly cluster state, and to reveal the range of the pump parameters suitable for constructing the cluster state, we use the van Loock -- Furusawa separability criterion \cite{VLF}, applying it to nullifiers of neighbouring nodes \cite{korolev2017}:
\BY
&&\<\delta(\h N_1)^2\>+\<\delta(\h N_2)^2\>>1\nonumber,\nn\\
&&\<\delta(\h N_2)^2\>+\<\delta(\h N_3)^2\>>1\label{NES},\\
&&\<\delta(\h N_3)^2\>+\<\delta(\h N_4)^2\>>1\nonumber.\nn
\EY

The violation of the inequalities (\ref{NES}) indicates that the state under consideration is a cluster state. Making the necessary substitutions, we find that the criterion (\ref{NES}) is violated for all related pairs of cluster nodes in a fairly wide range of pump parameters, as shown in Fig.\ref{Fig.5}.

It is seen from Fig.\ref{Fig.5} that when the pump parameter tends to a threshold value, $\mu \to\sqrt{\frac{2}{5}}$, for pairs of nodes at the edges the limit of the sum of spectral powers of fluctuations of the nullifiers is greater than the corresponding value for the pairs of nodes at the middle. This is primarily due to the non-perfect squeezing of supermodes $ \h S_2 $ and $ \h S_3 $.
\begin{figure}[h!]
\includegraphics[scale=0.68]{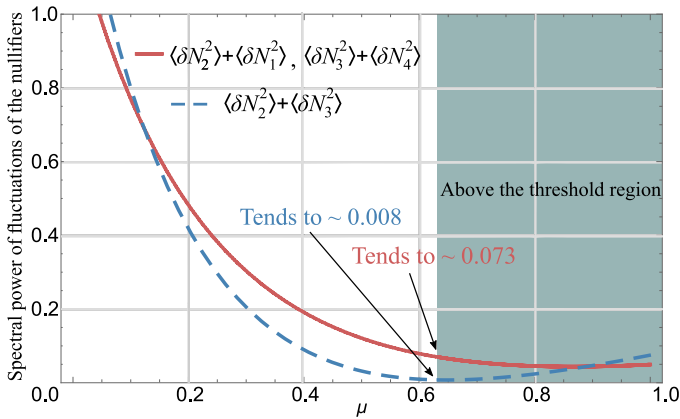}
\caption{The fulfillment of the van Look--Furusawa criterion for different values of the pump parameter $\mu$. }
\L{Fig.5}
\end{figure}

The state of the field is a four-mode linear cluster state in a wide range of $\mu$, starting at $\mu\approx0.065$. In this case, the smallest values of the variances of the nullifiers are achieved as we approach the threshold:
 \BY
 &&\lim_{\mu\to\sqrt{\frac{2}{5}}}\(\<\delta(\h N_4)^2\>\)=\lim_{\mu\to\sqrt{\frac{2}{5}}}\(\<\delta(\h N_1)^2\>\) \approx0.07,\\
 &&\lim_{\mu\to\sqrt{\frac{2}{5}}}\(\<\delta(\h N_3)^2\>\)=\lim_{\mu\to\sqrt{\frac{2}{5}}}\(\<\delta(\h N_2)^2\>\)\approx0.004.\;\;\;\;
 \EY
Recall that the pump parameter $\mu$ is related to the pump amplitude by the following relation:
\BY\mu=\displaystyle\frac{2\chi_{0,1}B_1}{\gamma}=\frac{2\chi_{0,1}B_{-1}}{\gamma}.\EY

Comparing our results with the works \cite{chinese, chinese2}, we can say that we managed to obtain a cluster state with a higher degree of entanglement between the nodes. In our case, the decrease in the spectral power of quadrature fluctuations of nullifiers reaches $ \sim-11$ dB, which significantly exceeds the values obtained for the cluster states in the cited works. This result is achieved through the usage of the supermode technique and the choice of the appropriate basis.
\section{Changes in the quantum properties of the system by varying the cavity configuration}
It is interesting to follow the changes in the quantum properties of the cluster state in dependence on the parameters of the cavity, such as the ratio of the waist width of the signal mode to the waist width of the pump mode, $r =w_s(0)/w_{pump}(0) $. Recall that estimates (41)-(42) of the cluster state given in the previous section performed under condition $r=\sqrt{2}$. This choice of $r$ provides a maximum of some coupling constants and this value was indicated by the authors of \cite{chinese} as the best for building a cluster state. Interested in comparing our results with those obtained earlier, we performed estimates for the same system parameters as in \cite{chinese}. However, looking at Fig.\ref{Fig.2b}, we can distinguish another specific value of the ratio of the waist width of beams: $r = 1$. This value is “suspicious” \; for a good result: with equal values\;of waist widths, the maximum of coupling constants of other modes involved in the process is ensured. Let us check how the resulting cluster state changes in this case.

Equations for the power spectra of fluctuations of the quadratures of the initial modes (\ref{quad}), as well as analytical expressions for the supermodes (\ref{s1})-(\ref{s5}), have been obtained in general form and can be directly used for further analysis. At the same time, the pump parameters and the ratio (\ref{pp}) between them have changed:
\BY \widetilde{\mu}^2=\frac{9}{8}\widetilde{\xi}^2. \EY
Here, $\widetilde{\mu}, \widetilde{\xi} $ are the new pump parameters for $r=1$. At the same time, the threshold value of the pump parameter is shifted. The condition on the below threshold regime of field generation is rewritten as $\displaystyle\widetilde{\mu}<\sqrt{\frac{9}{26}}$. Despite of the fact that the below threshold range of variation of parameters narrowed a little, the spectral powers of quadrature fluctuations in the supermodes $\h S_2, \h S_3$, as they approached the threshold, decreased in comparison to those calculated at $r = \sqrt{2}$:
\BY &&\lim_{\widetilde\mu\to\sqrt{\frac{9}{26}}}=\(\<|\delta \h X^S_2|^2\>\)=\lim_{\widetilde\mu\to\sqrt{\frac{9}{26}}}\(\<|\delta \h Y^S_3|^2\>\)\approx0.02.\;\;\;\;\;\;\;\;
\EY
That leads to changes in the limit values of the variances of the nullifiers:
\BY
&&\lim_{\widetilde\mu\to\sqrt{\frac{9}{26}}}\(\<\delta(\h N_4)^2\>\)=\lim_{\widetilde\mu\to\sqrt{\frac{9}{26}}}\(\<\delta(\h N_1)^2\>\) \approx0.02,\;\;\;\;\;\;\;\\
&&\lim_{\widetilde\mu\to\sqrt{\frac{9}{26}}}\(\<\delta(\h N_3)^2\>\)=\lim_{\widetilde\mu\to\sqrt{\frac{9}{26}}}\(\<\delta(\h N_2)^2\>\)\approx0.005.\nn
\EY
It can be noted that the new limit values of the variances of nullifiers $\h N_4, \h N_1$ have significantly decreased, which positively affects the cluster state entanglement, despite the small increase in the corresponding values for nullifiers $ \h N_2, \h N_3 $. Since the "quality" of the cluster state is determined by the sum combinations of variances of the nullifiers (\ref{NES}), we can confidently say about the improvement of cluster entanglement with $r=1$.

\begin{figure}[t!]
\includegraphics[scale=0.6]{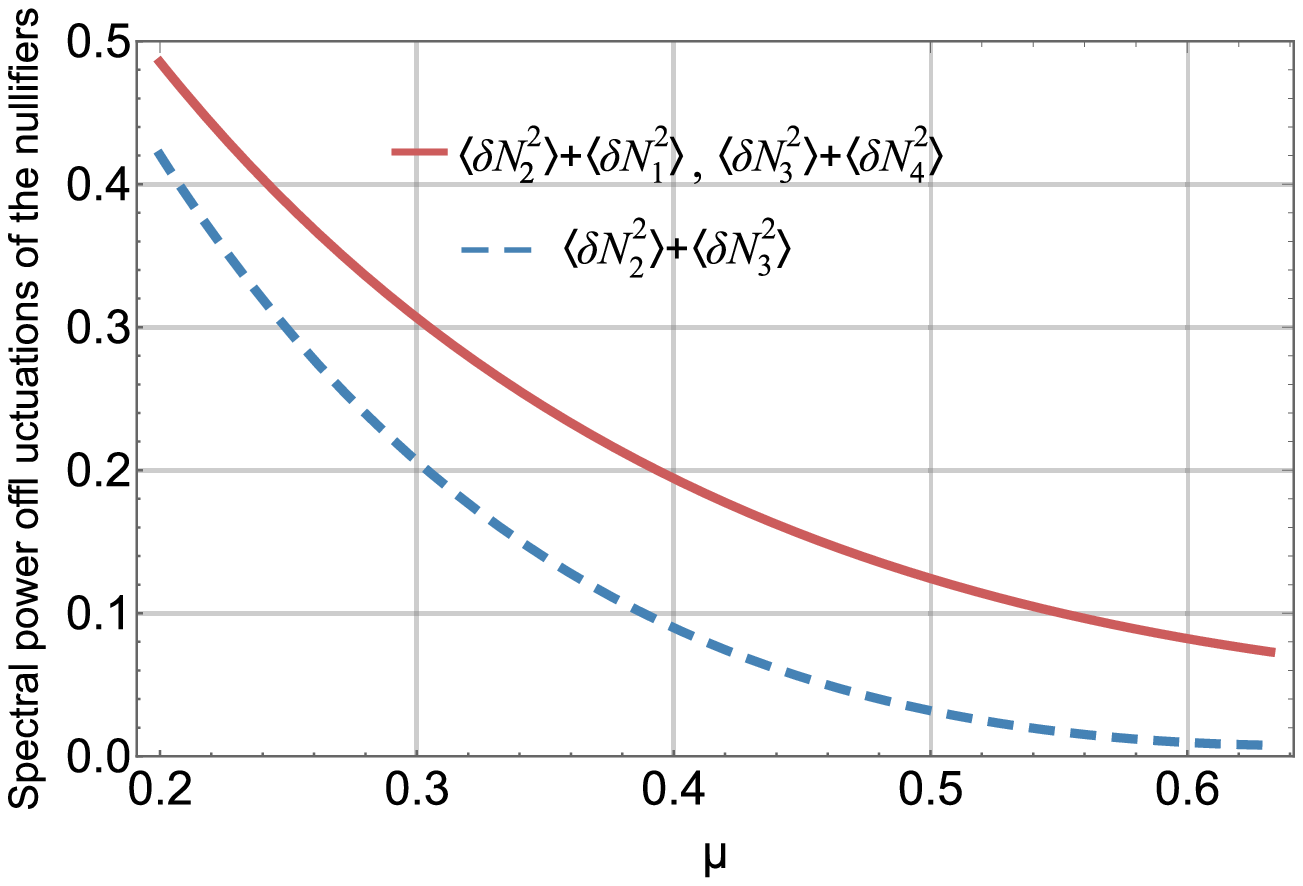}\\
\includegraphics[scale=0.601]{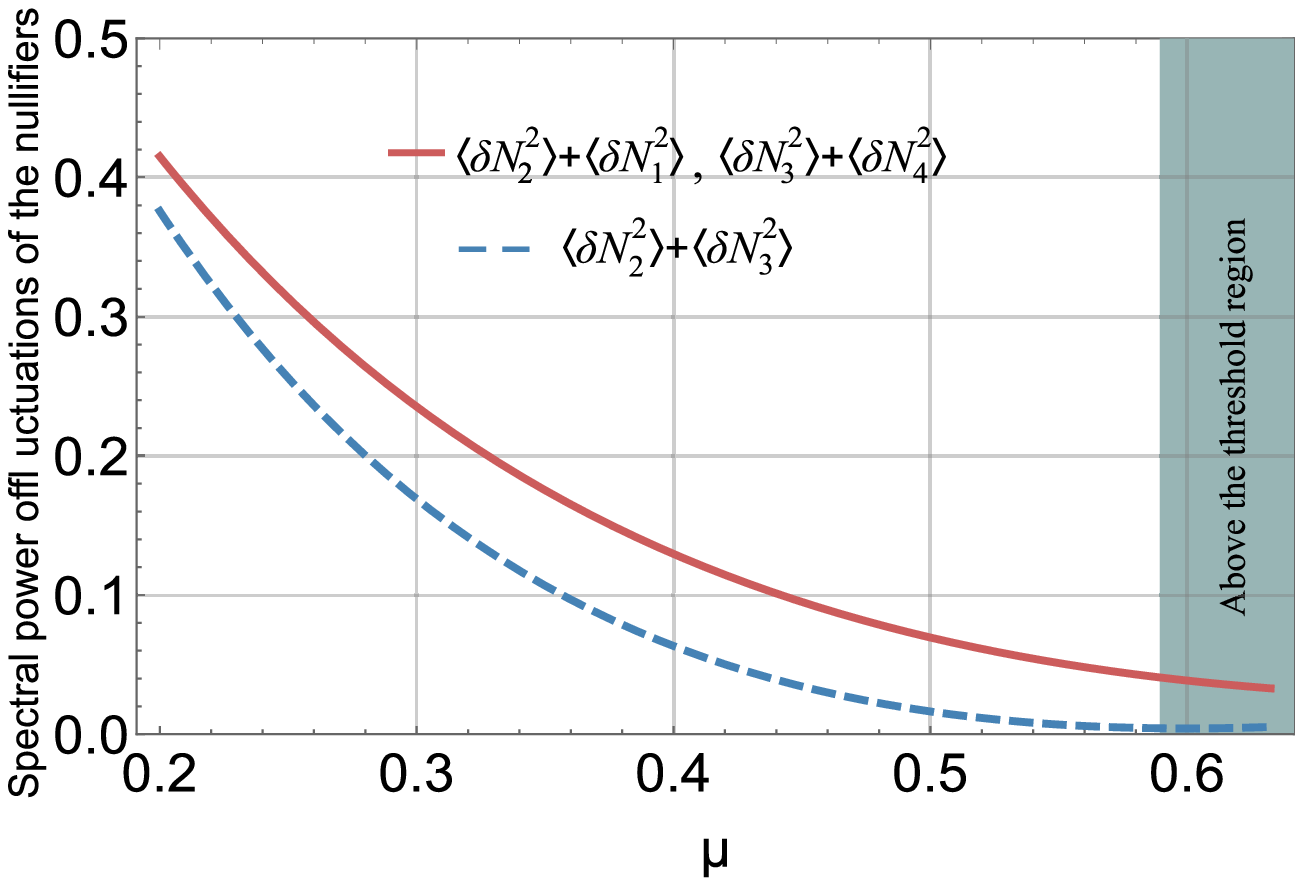}
\caption{Top — sums of variances of the nullifiers of neighboring nodes with the ratio of waist widths $r = w_s (0)/ w_p (0) =\sqrt{2}$. Bottom - the same values for the ratio $r = w_s (0) / w_p (0) = 1 $.}

\L{Fig.7}
\end{figure}

Discussing the reasons for this improvement of the quality of the cluster, two competing factors should be identified. On the one hand, the shift of the threshold leads to an increase in quadrature fluctuations in the modes with a higher degree of squeezing (see Fig.\ref{Fig.7}). On the other hand, the fluctuations themselves become smaller, which is especially noticeable in the worse squeezed modes. A decrease in values $\<\delta(\h N_4)^2\>,\<\delta(\h N_5)^2\>$ means that the effect of reducing quadrature fluctuations exceeds the effect of a threshold shift.

Thus, the choice of parameters does not affect the structure of the supermodes, but shifts the threshold value of the pump parameters and changes the spectral powers of the quadrature fluctuations of the supermodes. For the construction of a four-mode linear cluster, it is preferable to choose the equal values of waist widths of the pump and signal beams.
\section{Conclusion}
We have analysed the quantum properties of the field with an orbital angular momentum, generated in the process of spontaneous parametric down-conversion when the system is pumped by two Laguerre-Gaussian modes with OAM projections equal $1$ and $-1$.

This choice of pump is due primarily to the rich mode structure generated in such a system. If the angular momentum of the pump waves is set to $0$, then the OAM-degenerate process of generation of modes with zero orbital angular momentum will prevail over other possible non-degenerate processes (see Fig.\ref{Fig.2b}). If we pump the system with Laguerre-Gaussian modes of a higher order, then presumably instead of one multipartite quantum system, this will lead to the creation of several clusters of lower dimension. Thus, we stopped at the simplest pump that provides non-trivial system entanglement. More complex pump configurations require further analysis.

Treating of the $ n $-mode problem inevitably leads to the search for solutions of  $n$ differential equations, which mesh with each other, so it is necessary to limit the number of modes under consideration. This can be done, for example, experimentally, providing a fast relaxation of one of the modes, with the result that the system is divided into two subsystems, which can be solved independently.

There is an entanglement between modes with different OAM, however, this basis is not optimal for building a cluster state. To search for the optimal basis, we used the technique developed in \cite{Treps2014}, the method for revealing the supermodes, which are the eigenvectors of the coupling matrix $M$. It is necessary to emphasize the differences of our task: in the original paper, the authors considered a multimode system consisting of $10 ^ 6$ modes, that is, a quasi-continuous spectrum was subjected to analysis. First of all, this situation has limited the authors only by numerical analysis. In our case, supermodes can be found analytically. The analysis of the quasi-continuous spectrum is also connected with another important difficulty: although it was theoretically expected to get about 100 squeezed supermodes,  squeezing has been observed experimentally in only 6 of them \cite{SPOPO2014}.
 The authors attribute this effect to the fact that the measurement process implies a procedure of discretization of the spectrum and "cutting out" $\;$ of the parts of a comb, which fatally leads to energy losses.
In our case, the modes are discrete, so the measurement procedure is not associated with the spectrum discretization losses,which indicated above.

The calculations showed that of the five supermode only four can be used to build the cluster state since one supermode is in the vacuum state. Thus, despite the presence of quantum correlations between all five initial modes, the system has only four genuine quantum degrees of freedom.

The analysis of the region near the oscillation threshold is of the greatest interest. It turns out that the squeezing in some supermodes  is not perfect, which worsens the prospects for building a cluster state, but the estimation of the cluster state using the van Look--Furusawa criterion suggests that near the threshold the resulting multipartite entangled state is the cluster state in a fairly wide range of the pump power.

All the above reasoning can be generalized to any desired number of initial modes with a certain OAM. We concentrated on building a linear cluster state from four nodes because, as shown in \cite{nielsen2006, Furusava2010}, this number of degrees of freedom of the system is minimally necessary to perform the full set of single-mode and two-mode logical gates for the one-way quantum calculations.
If we include a larger number of the modes into consideration, the quadrature squeezing in the supermodes will deteriorate and, as a result, “degrade” the $ \; $ cluster state due to the same threshold limits for the pump power, which we discussed in Section IV.

This work was supported by RFBR (grants 18-02-00648a and 16-02-00180a). Scientific results were achieved during the implementation of the Program within the state support of the STI Center "Quantum Technologies".

\appendix
\section{Calculation of correlations of c-numerical stochastic noise sources}
Let us calculate the diffusion coefficients for c-number Langevin equations, taking into account that the equations for the variances of c-number quantities must coincide with the corresponding operator equations \cite{Davidovich}. The correlation function $\langle \h f_{0}(t),\h f^\dag_{0}(t^\prime) \rangle$ can be calculated from the system of Eqs. (\ref{eqn}):
\BY
&&\frac{d}{dt}\langle \h a_0 \h a^\dag_0\rangle=\langle \dot{\h a}_0\h a^\dag_0+ \h a_0\dot{\h a}^\dag_0\rangle=-\gamma\langle \h a_0 \h a_0^\dag\rangle +\\
&&+ 2\chi_{0,1}(B_1\langle \h a^\dag_{1}\h a^\dag_0\rangle +B_{-1}\langle \h a^\dag_{-1}\h a^\dag_0\rangle)-\gamma \langle\h a_0 \h a^\dag_0\rangle+\nonumber\\
&& + 2\chi_{0,1}\left( B_1\langle\h a_0\h a_{1}\rangle+B_{-1}\langle\h a_0 \h a_{-1}\rangle\right)+2D_{\h a_0 \h a_0^\dag}.\nonumber
\EY
We obtain the following equation after the normal ordering of the operators:
\BY
&&\frac{d}{dt}\langle \h a_0 \h a^\dag_0\rangle= + 2\chi_{0,1}(B_1\langle \h a^\dag_{1}\h a^\dag_0\rangle +B_{-1}\langle \h a^\dag_{-1}\h a^\dag_0\rangle)+\nonumber\\
&&+ 2\chi_{0,1}\left( B_1\langle\h a_0\h a_{1}\rangle+B_{-1}\langle\h a_0 \h a_{-1}\rangle\right) +2D_{\h a_0 \h a_0^\dag} -\nn\\
&&-2\gamma_a(1+\langle \h a^\dag_0 \h a_0\rangle).
\EY
On the other hand, from the corresponding c-number equations we can write:
\BY
&&\frac{d}{dt}\langle  a_0  a^*_0\rangle= 2\chi_{0,1}\left(B_1\langle  a^*_{1} a^*_0\rangle +B_{-1}\langle  a^*_{-1} a^*_0\rangle\right)+\nonumber\\
&&+ 2\chi_{0,1}\left( B_1\langle a_0 a_{1}\rangle+B_{-1}\langle a_0  a_{-1}\rangle\right) +2\EuScript{D}_{ a_0  a_0^*}-\nn\\
&& -2\gamma_a\langle a^*_0 a_0\rangle.
\EY
Given that $D_{\h a_0 \h a_0^\dag}=\gamma_a$, a new correlation function can be written:
$$
\langle  f_{0}(t), f^*_{0}(t^\prime)\rangle=2\EuScript{D}_{ a_0  a_0^*}\delta(t-t^\prime)=0.
$$
Other correlation functions are calculated in the same way. We write down only non-zero correlators for brevity:
\BY
\langle f_{0}(t) f_{\pm1}(t^\prime)\rangle = \gamma\mu\delta(t-t^\prime),\\
\langle  f_{{\pm1}}(t), f_{{\mp2}}(t^\prime)\rangle=\gamma\xi\delta(t-t^\prime).\nonumber
\EY
\section{Solution of system of the Heisenberg-Langevin equations and the calculation of the quadrature fluctuations}
To solve the system of Eqs. (\ref{qd}), we perform a Fourier transform:
 \BY
\binom{x^\Omega_{i}}{y^\Omega_{i}}= \frac{1}{2\pi}\int\limits_{-\infty}^{\infty}dt\binom{x_{i}(t)}{y_{i}(t)}e^{-i\Omega t},\nonumber\\
\binom{f_{i}^{\prime\Omega}}{f_{i}^{\prime\prime\Omega}}=\frac{1}{2\pi}\int\limits_{-\infty}^{\infty}dt\binom{f_{i}^\prime}{f_{i}^{\prime\prime}}e^{-i\Omega t}.\nonumber
\EY
As a result, we obtain a system of linear algebraic equations, and we will be interested in the solution of this system at zero frequency.

The solution of the system (\ref{ft}) has the form (hereinafter the quadrature components of Langevin noise sources have the superscript 0 omitted):
\BY
&&x^0_{0}=\frac{1}{\gamma(1-2\mu^2-\xi^2)}[(1-\xi^2)f^{\prime}_{0} + \mu(f^{\prime}_{1}+f^{\prime}_{-1}) + \nn\\ &&+\mu\xi(f^{\prime}_{2}+f^{\prime}_{-2})],\nonumber\\
&&x^0_{\pm1}=\frac{1}{\gamma(1-2\mu^2-\xi^2)}[\mu f^{\prime}_{0} + f^{\prime}_{\pm1} + \xi f^{\prime}_{\mp2}+\nn\\
&&+\frac{\mu^2}{1-\xi^2}\left(f^{\prime}_{\mp1}-f^{\prime}_{\pm1} + \xi(f^{\prime}_{\pm2}-f^{\prime}_{\mp2})\right)],
\EY
\BY
&&x^0_{\pm2}=\frac{1}{\gamma(1-2\mu^2-\xi^2)}[\mu\xi f^{\prime}_{0} + \xi f^{\prime}_{\mp1} + (1-2\mu^2)f^{\prime}_{\pm2}+\nn\\
&&+\frac{\mu^2\xi}{1-\xi^2}\left(f^{\prime}_{\pm1}-f^{\prime}_{\mp1} + \xi(f^{\prime}_{\mp2}-f^{\prime}_{\pm2})\right)].\nonumber
\EY

The formulas become more compact when moving to new variables $x_0^0=\widetilde{x}_0$, $\widetilde{ x}_1= \frac{1}{\sqrt{2}}(x^0_{1}+x^0_{-1}) $, $\widetilde{x}_2= \frac{1}{\sqrt{2}}(x^0_{2}+x^0_{-2})$, $\widetilde{ x}^-_1= \frac{1}{\sqrt{2}}(x^0_{1}-x^0_{-1})$ and $\widetilde{x}^-_2= \frac{1}{\sqrt{2}}(x^0_{2}-x^0_{-2})$. New stochastic noise sources  $\widetilde{f^{\prime}_{i}}, \widetilde{f^{\prime}_{i}}^-$ are related to the old ones respectively.
\BY
&&\widetilde{x}_{0}=\frac{1}{\gamma(1-2\mu^2-\xi^2)}\left[(1-\xi^2)f^{\prime}_{0} + \sqrt{2}\mu\widetilde{f^{\prime}_{1}} + \sqrt{2}\mu\xi\widetilde{f^{\prime}_{2}}\right],\nonumber\\
&&\widetilde{x}_{1}=\frac{1}{\gamma(1-2\mu^2-\xi^2)}\left[\sqrt{2}\mu f^{\prime}_{0} + \widetilde{f^{\prime}_{1}} + \xi \widetilde{f^{\prime}_{2}}\right],\nonumber\\
&&\widetilde{x}_{2}=\frac{1}{\gamma(1-2\mu^2-\xi^2)}\left[\sqrt{2}\mu\xi f^{\prime}_{0} + \xi \widetilde{f^{\prime}_{1}} + (1-2\mu^2)\widetilde{f^{\prime}_{2}}\right],\nonumber\\
&&\widetilde{x}^-_{1}=\frac{1}{\gamma(1-\xi^2)}\left[\widetilde{f^{\prime}_{1}}^- - \xi \widetilde{f^{\prime}_{2}}^-\right],\nonumber\\
&&\widetilde{x}^-_{2}=\frac{1}{\gamma(1-\xi^2)}\left[\widetilde{f^{\prime}_{2}}^- - \xi \widetilde{f^{\prime}_{1}}^-\right].\nonumber
\EY

Correlators of the redefined Langevin's noise sources are given by:
\BY
&&\langle  f^\prime_{{0}},\widetilde{f^{\prime}_{1}}\rangle=\frac{\gamma\mu}{\sqrt{2}},\;\;\;\;
\langle \widetilde{f^{\prime}_{1}}, \widetilde{f^{\prime}_{2}}\rangle=\frac{\gamma\xi}{2},\nn\\
&&\langle \widetilde{f^{\prime}_{1}}^-, \widetilde{f^{\prime}_{2}}^-\rangle=-\frac{\gamma\xi}{2}.
\EY

Let us calculate the power spectra of fluctuations of the quadrature components and the cross-correlators:
\BY
&&\langle |\delta \widetilde{x}_0|^2\rangle =\frac{2\mu^2}{\gamma(1-\xi^2-2\mu^2)^2}=\frac{\xi^2}{2\mu^2}\langle |\delta \widetilde{x}_{2}|^2\rangle, \nn\\
&&\langle |\delta \widetilde{x}_{1}|^2\rangle =\langle |\delta x_0|^2\rangle +\langle |\delta \widetilde{x}_{2}|^2\rangle,  \nn\\
&&\langle \delta \widetilde{x}_0\delta \widetilde{x}_{1}\rangle=\frac{\mu(1+ 2\mu^2 + \xi^2)}{\sqrt{2}\gamma(1-\xi^2-2\mu^2)^2}=\frac{\xi}{\sqrt{2}\mu}
\langle \delta \widetilde{x}_{1}\delta \widetilde{x}_{2}\rangle,\nn\\
&&\langle \delta \widetilde{x}_0\delta
\widetilde{x}_{2}\rangle=\frac{\sqrt{2}\mu\xi}{\gamma(1-\xi^2-2\mu^2)^2},\nonumber\\
&&\langle |\delta \widetilde{x}^-_{1}|^2\rangle =\frac{\xi^2}{\gamma(1-\xi^2)^2}=\langle |\delta \widetilde{x}^-_{2}|^2\rangle,\\
&&\langle \delta \widetilde{x}^-_{1}\delta \widetilde{x}^-_{2}\rangle=-\frac{\xi(1+ \xi^2)}{2\gamma(1-\xi^2)^2}.\nonumber
\EY
Since the equations for the differential modes $\widetilde{x}^-_{i}, (i=0, \pm1, \pm2)$ form a closed system, their power spectra and cross-correlations do not depend explicitly on the parameter $\mu$.

Carrying out all of the same steps to calculate power spectra of fluctuations of the $y$-quadratures, we get:
\BY
&&\langle |\delta \widetilde{y}_0|^2\rangle =\langle |\delta \widetilde{x}_0|^2\rangle,\;
\langle |\delta \widetilde{y}_{1}|^2\rangle =\langle |\delta \widetilde{x}_{1}|^2\rangle ,\;\langle |\delta \widetilde{y}_{2}|^2\rangle =\<|\delta \widetilde{x}_{2}|^2\>,\nn\\
&&\langle \delta \widetilde{y}_0\delta \widetilde{y}_{1}\rangle=-\langle \delta \widetilde{x}_0\delta \widetilde{x}_{1}\rangle,\;\langle \delta \widetilde{y}_{1}\delta \widetilde{y}_{2}\rangle=-\langle \delta \widetilde{x}_{1}\delta \widetilde{x}_{2}\rangle,\nn\\
&&\langle \delta \widetilde{y}_0\delta \widetilde{y}_{2}\rangle=\langle \delta \widetilde{x}_0\delta \widetilde{x}_{2}\rangle,\nn\\
&&\langle |\delta \widetilde{y}^-_{1}|^2\rangle =\langle |\delta \widetilde{x}^-_{1}|^2\rangle ,\langle |\delta \widetilde{y}^-_{2}|^2\rangle =\<|\delta \widetilde{x}^-_{2}|^2\>,\nn\\
&&\langle \delta \widetilde{y}^-_{1}\delta \widetilde{y}^-_{2}\rangle=-\langle \delta \widetilde{x}^-_{1}\delta \widetilde{x}^-_{2}\rangle.
\EY
\end{document}